\documentclass[acm,final]{paper}

\begin{document}

\title{\sys: A Peer-to-Peer Network with Verifiable Causality}

\ifauthor
\author{Michael Hu Yiqing}
\affiliation{
    \institution{National University of Singapore}
}
\email{hmichael@nus.edu.sg}
\author{Guangda Sun}
\affiliation{
    \institution{National University of Singapore}
}
\email{sung@comp.nus.edu.sg}
\author{Arun Fu}
\affiliation{
    \institution{Advaita Labs}
}
\email{arun@advaita.xyz}
\author{Akasha Zhu}
\affiliation{
    \institution{Advaita Labs}
}
\email{akasha@advaita.xyz}
\author{Jialin Li}
\affiliation{
    \institution{National University of Singapore}
}
\email{lijl@comp.nus.edu.sg}
\else
\author{In Submission}
\fi

\begin{abstract}
Logical clocks are a fundamental tool to establish causal ordering of events in a distributed system.
They have been used as the building block in weakly consistent storage systems, causally ordered broadcast, distributed snapshots, deadlock detection, and distributed system debugging.
However, prior logical clock constructs fail to work in a permissionless setting with Byzantine participants.
In this work, we introduce \sys, a novel logical clock system that targets an open and decentralized network.
\sys introduces a new logical clock construct, the Decaying Onion Bloom Clock (\clk), that scales independently to the size of the network.
To tolerate Byzantine behaviors, \sys leverages non-uniform incrementally verifiable computation (IVC) to efficiently prove and verify the construction of \clk clocks.
We have applied \sys to build two decentralized applications, a weakly consistent key-value store and an anti-censorship social network, demonstrating the power of scalable, verifiable causality in a decentralized network. \end{abstract}

\maketitle

\section{Introduction}
\label{sec:intro}
The ordering of events is a fundamental concept in distributed systems.
In state machine replication systems~\cite{smr,paxos,vr,raft}, the set of replicas needs to agree on the order of operations in the log;
shards in a distributed database~\cite{spanner,megastore,calvin} are tasked to execute distributed transactions in a consistent partial order;
for mutual exclusion of shared resources, participants in a distributed system have to agree on the order of acquiring locks~\cite{chubby,lockmanager,netlock};
in a distributed storage system~\cite{gfs,ceph,bigtable,dynamo,pegasus}, servers apply a consistent order of mutations to storage objects.

It is well-known that perfectly synchronized clocks do not exist in realistic distributed systems, due to clock drift and relativity.
Ordering events using physical clock timestamps is therefore not reliable and can lead to anomalies.
Logical clocks~\cite{lamport-clock,vector-clock}, on the other hand, offer a solution to order events in a distributed system without relying on physical time.
Logical clocks are consistent with \emph{logical causality}, \ie, if event $a$ can causally influence event $b$, then the logical clock of $a$ is prior to that of $b$.
Unlike physical time ordering, causality in a distributed system is only a partial order, as there exists events which do not causally influence each other.
Many forms of logical clocks have been proposed in the literature~\cite{lamport-clock,vector-clock,bloom-clock,plausible-clock,tree-clock}, though not all of them can be used to deduce causality between events.
For instance, even if an event has a smaller Lamport clock~\cite{lamport-clock} than another, the two events can still be logically concurrent.

Existing logical clock constructs, however, fall short in an open, decentralized network~\cite{bitcoin,ethereum}.
In these networks, any participant can join or leave the system at any time.
Such dynamic environment presents deep scalability challenges for vector-based logical clocks~\cite{vector-clock,plausible-clock,tree-clock}.
More critically, prior systems assume all participants in the system faithfully follow the protocol to update and propagate their clocks.
In a decentralized network, Byzantine~\cite{byzantine} behaviors, where a participant can deviate arbitrarily from the specified protocol, are common.
Unfortunately, existing logical clock constructs are not Byzantine-fault tolerant.
By not following the clock protocol, a single Byzantine participant can easily compromise the clock's causality guarantees, \ie, logical clocks may imply erroneous causality between events.
Such adversaries can render the entire clock construct pointless.

In this work, we address the above shortcomings by proposing a new logical clock system, \sys.
\sys targets a permissionless network with possible Byzantine participants.
Similar to prior logical clocks, \sys can be used to infer causality in the network.
\sys, however, only concerns causal dependency between object states, \ie, an object state is the result of a series of mutations from another object state.
In return, \sys always infers \emph{true causality}, unlike the possible causality implied by prior approaches.
To handle dynamic membership, \sys introduces Decaying Onion Bloom Clock (\clk), a novel construct based on Bloom clocks~\cite{bloom-clock}.
\clk is agnostic to the identity and the number of the participants in the network.
It achieves this property by applying Bloom filters to only record the state transition history.
To maintain low false positive rate even for arbitrarily long causal histories, \clk uses layers of Bloom filters, a construct inspired by log-structured merge-tree~\cite{lsmtree}.
Recent transitions are stored in the top layer filters;
when a layer is filled up, its filters are merged and pushed to the next layer.
\clk therefore offers accurate causality inference for recent histories, while its accuracy gracefully degrades for the distant past.

To tolerate Byzantine participants, \sys builds upon recent advances in verifiable computation (VC).
Specifically, \sys applies non-uniform incrementally verifiable computation (IVC)~\cite{ivc}, a proof system that uses recursive Succinct Non-interactive Argument of Knowledge (SNARKs)~\cite{snarks}.
When mutating an object in \sys, the initiating node generates a succinct proof that demonstrates the validity of both the state transition and the \clk clock update.
The proof is attached to the object when disseminating the object in the network.
A receiver verifies the attached proof before accepting the object.
Using IVC, a node can incrementally mutate any verified object, and efficiently generate a succinct proof for the entire causal history of the object.
Moreover, both prover time and verifier time are independent of the length of the causal history.
As each node may apply arbitrary state mutation function to an object, \sys uses a variant of IVC called non-uniform IVC~\cite{Kothapalli2022SuperNovaPU} to address the rigidity of the original IVC construct.

We have built two decentralized systems atop \sys to demonstrate the power of verifiable causality.
The first system is a weakly consistent data store \kstore.
\kstore provides eventual consistency~\cite{bayou} even in the presence of strong adversaries.
It leverages \sys to track versioned histories of stored data and to effectively merge conflicting versions.
It relies on provable causal history to avoid lost updates or inconsistencies cause by Byzantine behaviors.
\sys is more scalable, available, and provides faster query latency than existing BFT systems.
The second system, \ksn, is an anti-censorship decentralized social network.
Beyond the benefits already provided in \kstore, \sys enables \ksn to effectively eliminate censorship attacks, a major challenge in other social applications.
Using verifiable causality, clients in \ksn can enforce propagation and visibility of posted content, and generate proof-of-censorship when censorship attacks are launched.
 
\section{Background}
\label{sec:background}
This section covers background information on two main topics: causality of events in a distributed system, and verifiable computation.

\subsection{Causality of Events in Distributed Systems}
\label{sec:background:causality}

The seminal work by Lamport~\cite{lamport-clock} introduces a \emph{happens-before} relationship that defines the possible causality between events in a distributed system.
Specifically, let $\prec$ be a binary relation between pairs of events in an execution of a distributed system.
$e_1 \prec e_2$ if event $e_1$ may influence event $e_2$, or equivalently, $e_2$ is causally dependent on $e_1$.
$\prec$ is a strict partial order, \ie, it is \emph{irreflexive}, \emph{asymmetric}, and \emph{transitive}.
Being a partial order, not all pairs of events are causally dependent.
If neither $e_1 \prec e_2$ nor $e_2 \prec e_1$, $e_1$ and $e_2$ are defined to be logically concurrent (represented as $e_1 \parallel e_2$).
Without perfectly synchronized physical clocks, it is impossible to determine which of $e_1$ or $e_2$ happens first if $e_1 \parallel e_2$.

Events in an execution are categorized into three general types:
\begin{itemize}
    \item Local events on a node (\elocal).
    They are any event happen on a node that does not involve messages.
    \item Message send event (\esend).
    A source node sends an unicast message to a destination node.
    Broadcasts or multicasts are equivalent to a set of unicast messages.
    \item Message receive event (\erecv).
    For each \esend, there is a corresponding message receive event on the destination node if the message is successfully delivered.
\end{itemize}

The happens-before relation $\prec$ on events in an execution obeys the following rules:
\begin{itemize}
    \item \el{1} $\prec$ \el{2} if both events happen on the same node and \el{1} happens before \el{2} in the local sequential event order.
    \item \esend $\prec$ \erecv if \esend and \erecv are the corresponding message send and receive pair.
\end{itemize}

\subsection{Logical Clocks}
\label{sec:background:lc}

Logical clocks can be used to determine the happens-before relation defined in \autoref{sec:background:causality}.
One instance of logical clocks is the Lamport clock~\cite{lamport-clock}.
Using Lamport clock, each node $n_i$ in the system maintains a local clock $c_i$, represented as a natural number.
The rules to update the clocks are:
\begin{itemize}
    \item Upon a local event \el{i}, $n_i$ increments its local clock.
    \item When $n_i$ sends a message, $n_i$ increments its local clock, and attaches the local clock value in the message.
    \item When $n_i$ receives a message with a clock value $c_m$, it sets its local clock to $max(c_i, c_m) + 1$.
\end{itemize}

The logical time of an event $e$, represented as $c_e$, is the local clock value after the clock update.
Lamport clock guarantees the following property: If $e_1 \prec e_2$, then $c_{e_1} < c_{e_2}$.
However, the inverse is not true, \ie, $c_{e_1} < c_{e_2}$ does not imply that $e_1 \prec e_2$.
To be more precise, if $c_{e_1} < c_{e_2}$, either $e_1 \prec e_2$ or $e_1 || e_2$, but not $e_2 \prec e_1$.
For use cases that require accurate causality inference, Lamport clock falls short.

Vector clock addresses this shortcoming of Lamport clock.
As the name suggested, a vector clock, $v$, consists of a vector of natural numbers.
Cardinality of a vector clock equals the size of the system.
Each node is assigned a unique index in the vector clock.
We use $v[i]$ to denote the $i$th number in a vector clock $v$.
The rules to update the vector clocks are:
\begin{itemize}
    \item Upon a local event \el{i}, $n_i$ increments $v_i[i]$.
    \item When $n_i$ sends a message, $n_i$ increments $v_i[i]$, and attaches $v_i$ in the message.
    \item When $n_i$ receives a message with a clock $v_m$, it sets $v_i$ to $v_i'$, where $\forall p \in [0..S), v_i'[p] = max(v_i[p], v_m[p])$, $S$ is the size of the system.
    $n_i$ then increments $v_i[i]$.
\end{itemize}

$v_i < v_j$ if and only if $\forall p \in [0..S), v_i[p] \le v_j[p]$ and $\exists p \in [0..S), v_i[p] < v_j[p]$.
By definition, there exists $v_i$ and $v_j$ such that neither $v_i < v_j$ nor $v_j < v_i$, \ie, $<$ is a partial order on the set of vector clocks.
Vector clock guarantees the following stronger property: $e_1 \prec e_2$ if and only if $v_{e_1} < v_{e_2}$.

\subsection{Verifiable Computation}

In systems where identities cannot be trusted (our target deployment model), publicly verifiable proofs are required to verify the claims made by the participants.
More concretely, if a node claims that the output of applying a certain function $\mu$ on input $x$ is $y$, the naive way to verify such a statement would be to re-execute the operation and compare the outputs.
Such an approach might not be viable when the verifier does not have enough computational resources to execute the function.
For instance, the current bitcoin blockchain is approximately 450 GB in size.
If a user wants to verify the latest block, he can either:

\begin{enumerate}
    \item Trust the person who provided the latest block to him (this is extremely unwise).
    \item Verify the whole chain himself from the genesis block; This takes a lot of compute time, storage space, and network bandwidth.
\end{enumerate}

An argument system is a cryptographic construct to achieve verifiable computation without trusting the entity performing the computation.
The goals of an argument system are quite simple.
For a given statement $\mu(x) \stackrel{?}{\rightarrow} y$, it produces an accompanying proof $\pi$.
This proof can be verified publicly to assert that the statement is true with all but a negligible probability.
More concretely, a prover $P$ wishes to convince a verifier $V$ that it knows some witness statement $w$ such that, for some public statement $x$ and arithmetic circuit $C$, $C(x,w)\rightarrow y$.

\paragraph{Properties of an argument system.}
There are two properties an argument system must satisfy:

\begin{enumerate}
    \item \textbf{Completeness:} A valid proof will always be accepted by a valid verifier.
    \item \textbf{(Knowledge) Soundness:} If a prover attempts to generate a proof without a valid witness, this proof will only be accepted by the verifier with a negligible probability.
\end{enumerate}

There are many types of argument system.
One of the most commonly used argument systems is Succinct Non-interactive Arguments of Knowledge (SNARK).
As the name suggested, using a SNARK, the verifier requires no further interaction with the prover, other than receiving the proof, when verifying;
the proof itself is also short, while the time to verify is fast (at most logarithmic to the circuit size).
If the witness $w$ cannot be derived by the verifier with sufficient probability, then the SNARK is also considered \emph{zero-knowledge} (zk-SNARK).
\sys does not require the zero-knowledge property, so we omit the details of zk-SNARKs.

\paragraph{Recursive proof systems.}

SNARKs are useful in many settings, e.g., cloud computing, as it allows a verifier to validate computationally expensive function executions in a fraction of the time to run it.
However, in some distributed computing scenarios, simply verifying a single execution is not sufficient.
Instead, we wish to verify a particular non-deterministic chain of executions.
Naively applying any off-the-shelf general circuit SNARK for every step will result in proofs and verification times that grows linearly.
In a highly evolving and volatile system, these metrics are unacceptable.

There has been some recent development in verifiable computation that has the potential to address the above challenges.
One particularly promising technique is recursive proof system.
In a recursive proof system, the prover recursively proves the correct execution of incremental computations.
Such technique can be applied to realize incrementally verifiable computation (IVC).
In IVC, in each step of the computation, the prover takes the output and proof of the previous step, and produces an output and proof for the next step.
A verifier only needs to verify the proof of a single step to ensure correct execution of the entire computation from genesis.
Critically, both prover and verifier time are independent of the length of the computation.
There exists quite a few constructions for recursive proofs in the wild, ranging from constructs like Halo\cite{Bowe2020RecursivePC} to PCD\cite{pcd}.
We envision more and more efficient constructs will be developed eventually.

\section{\sys}
\label{sec:network}
We first define the high-level model and properties of \sys.
The system consists of a set of nodes ($n_1, n_2, \dots$).
Nodes can create, destroy, and mutate \emph{objects}.
They can also send objects to other nodes in the system.
Besides \code{send} and \code{recv}, we define generic \code{create} and \code{mutate} functions:

\begin{itemize}
    \item ${create()} \rightarrow o$: the \code{create} function generates an object.
    \item ${mutate(o_1, o_2, \dots)} \rightarrow o'$: the \code{mutate} function takes a list of objects $o_1, o_2, \dots$ and generates a new object $o'$.
\end{itemize}

Unlike prior work~\cite{lamport-clock}, \sys concerns the causality of objects, not causality of events.
We similarly define a binary relation $\prec$ on the set of objects in an execution of a distributed system.
$\prec$ denotes the causal relationship between any two objects, \ie, $o_1 \prec o_2$ if and only if $o_2$ is causally dependent on $o_1$.
Object causality in $\sys$ is defined as follows:

\begin{itemize}
    \item If an object $o$ is generated from \code{create}, $o$ is not causally dependent on any other object in the system, \ie, $\forall o' \in O, o' \nprec o$, where $O$ is all objects ever generated in the execution.
    \item If an object $o$ is generated from $mutate(o_1, o_2, \dots)$, $o$ is causally dependent on $o_1, o_2, \dots$, \ie, $o_1 \prec o, o_2 \prec o, \dots$
\end{itemize}

We note that the causality definition in \sys is stronger than those in prior work~\cite{lamport-clock}.
Instead of ``possible influence'', $\prec$ implies definite causal relationship between two objects\footnote{We assume \code{mutate} implies definite causality. That is, if $mutate(o_1, o_2, \dots) \rightarrow o$, then $o$ is causally dependent on $o_1, o_2, \dots$.}.
More formally, if $o_i \prec o_j$, there exist a sequence of \code{mutate} invocations such that $mutate(o_i, \dots) \rightarrow o_1$, $mutate(o_1, \dots) \rightarrow o_2$, \dots, $mutate(o_n, \dots) \rightarrow o_j$.

\sys provides the following \emph{causality inference} guarantee:

\begin{theorem}
For any two objects $o_i$ and $o_j$ which are generated in an execution of a distributed system, \sys can deduce the causality relationship between the two objects, \ie, \sys can correctly output $o_i \prec o_j$, $o_j \prec o_i$, or $o_i \parallel o_j$.
\end{theorem}
 
\section{Design and Implementation}
\label{sec:design}

This section details the concrete design of \sys using a novel logical clock construct coupled with verifiable computation.

\subsection{System Setting}
\label{sec:design:sys_settings}
\mh{TODO: double check some notations, esp after re-writing the verifiable computation part}

Suppose each object is defined as a tuple $o_i=(s_i,C_i)$, where $s_i \in \mathcal{S}$ is a state permutation in the set of all possible states $\mathcal{S}$.
$C_i$ is a logical clock construct.
This indicates that for two objects $o_1=(s_1,c_x), o_2 = (s_1,c_y)$ with homogeneous states, they are still considered unique $o_1\neq o_2$ if the clock values are distinct.

\sys attributes two objects with the exact same state and clock values as identical.
This, however, is not necessarily true.
This is inevitable and might lead to certain false positives in casualty evaluations.

There exists a dynamic set of objects $O$, which initially only includes the genesis object $o_0=(s_0,C_0,d)$, where $s_0$ is the genesis state.
There also exists a family of \code{mutate} functions $ M=\{\mu_1,\cdots\}$.
For a new object $o_i$ to be created, a \code{mutate} function must be operated on an existing object:

\begin{gather*}
    \exists o_{i-1} \in O, \exists \mu \in M: \mu(o_{i-1}) \rightarrow o_i, o_i \in O
\end{gather*}

Therefore, $d$ for an object denotes the depth, or number of \code{mutate} functions applied onto $o_0$ to derive the object.  
\subsection{Logical clock Constructs}
\label{sec:design:logical_clock_constructs}

In this section, we will explore a few existing clock constructs and evaluate its feasibility of use in \sys.

\subsubsection{Vector clocks}

Vector clocks~\cite{vector-clock} have long been the standard to determine causality in systems.
However, as briefly mentioned in \autoref*{sec:network}, vector clocks do not necessarily draw accurate object causalities.
Instead, vector clocks simply provide a temporal order and use that to infer possible causality;
This will lead to great degrees of false positives.

We illustrate this with a simple example; Suppose we have two nodes $A$ and $B$:

\begin{enumerate}
    \item $A$ and $B$ start with the vector clock $[0,0]$
    \item $A$ creates a new state, tags it with the clock value $[1,0]$ and sends it to $B$.
    \item $B$ receives this new state and updates its clock to $[1,1]$.
    \item Suppose $B$ now creates a new state by applying \code{mutate} to the genesis state instead of $A$'s newly produced state.
    $B$'s new clock value will be $[1,2]$.
\end{enumerate}

The issue is that although $B$'s new state does not depend on $A$'s state, its clock implies such.
This is because the vector clock only captures the temporal order of states/objects produced by the nodes, and therefore probable causality;
Furthermore, we conjecture this probability to be impossible to be derived.

Instead, we have to provide a way for $B$ to generate distinct clocks based on its decision (to either \code{mutate} the genesis object or $A$'s object).
This requires the decoupling of identities and clocks.
Instead, each object is now tied to a logical clock that is used to infer its plausible relationship with another object.

\subsubsection{Counting bloom clocks}
\label{sec:design:cbc}

The Bloom clock ($BC$)~\cite{bloom-clock} is a logical clock construct that can be used to probabilistically determine causality between objects.
The $BC$ is based on the counting Bloom filter~\cite{bloom-filter}, and can be defined as a vector of $n$ integers $[c_1,\cdots,c_n]$.

In the context of \sys, when operating on an existing object $\mu(s_i,C_i,d) \rightarrow (s_{i+1},C_{i+1},d+1)$, the $BC$ protocol uses a family of $m$ cryptographically secure hash functions $h_1,\cdots,h_m$ that produces $m$ indices $h_1(s_{i+1}), \cdots, h_m(s_{i+1})$.
Each index is then mapped and incremented on $C_i$ to produce $C_{i+1}$.

When comparing two objects, $o_x = (s_x,C_x,d_x)$ and $o_y = (s_y,C_y,d_y)$, there are 3 possible scenarios:
\begin{enumerate}
    \item
    $\forall c_{xi} \in C_x, c_{yi} \in C_y, \exists  c_{xj} \in C_x, c_{yj} \in C_y : c_{xi} \geq c_{yi} \land c_{xj} > c_{yj} \land d_x>d_y \Rightarrow (s_x,C_x) \succ  (s_y,C_y)$.
    \item
    $\forall c_{yi} \in C_y, c_{xi} \in C_x: c_{yi} \geq c_{xi} \land d_y \geq d_x \Rightarrow (s_y,C_y) \succeq (s_x,C_x) $
    \item $(s_y,C_y,d_y)$ and $(s_x,C_x,d_x)$ are concurrent.
\end{enumerate}

The $BC$ might postulate parenthood when in fact the objects are concurrent.
This is due to possible hash collisions into the limited vector of size $n$.

The fundamental benefit of $BC$'s is its inherent agnosticism towards identities in the system;
It can potentially be utilized by an unbounded number of nodes.
This makes it suitable to be utilized in highly decentralized and permissionless settings.

However, $BC$ suffers from two main issues:

\paragraph{\textbf{Issue 1: limited lifespan.}}
Eventually, a $BC$ will increment to a point in which comparisons between two objects will always lead to a false positive.
That is to say the $BC$ can only hold a limited amount of objects before its utility is diminished.
Therefore, to maintain its utility, we have to limit the number of objects a $BC$ can hold.
A naive reset might work, however it is crude.
Comparisons between resets are impossible.
Additionally, some synchrony might be required to derive consensus on the state of the clock before resets.

A simple fix would be to have a $BC$ be represented by $k$ bloom filters $(BF)$ of size $n$.
These $k$ bloom filters is a sliding window on the history of objects, where objects outside these $k$ bloom filters are forgotten.

The intuition is that after $k$ \code{mutate}s, there is little to be gained from comparing a descendant state that is so far removed from its ancestor.
Following that philosophy, more priority should be put into recent states (relative to the current state), and less priority to distant states.
Especially in the $BC$ protocol, a similar comparison will likely return a false positive.

\paragraph{\textbf{Issue 2: multi-parent problem.}}

In the $BC$ protocol, all prior object clocks $C_i$ are compressed into a single vector.
However, this also means that similar objects (by means of indices after hashing) represented in the clock will potentially lead to false positives.
The simple fix represented above resolves this to some extent.
Since each $BC$ is now represented by $k$ bloom filters, we can analyze the state at each depth of \texttt{mutation}.
This removes some false positives that previously existed in the original $BC$.

\begin{figure}[h!]
    \centering
    \includegraphics[width=.5\textwidth]{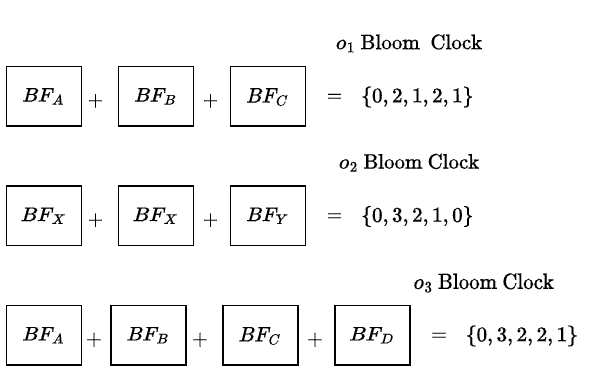}
    \caption{How adding some history can be used to eliminate some false positives: Without the history, $o_2$ will be incorrectly determined as the ancestor to $o_3$. }
    \label{fig:naive_false_parent}
\end{figure}

As illustrated with \autoref*{fig:naive_false_parent}, simply utilizing the bloom clock will lead to the false conclusion that $o_3$ is causally dependant on $o_2$;This is due to the coincidental hash collisions. However, if a history of $BF$s are listed, it can be referenced and compared to eliminate $o_2$ as a false parent of $o_3$.

\subsection{Decaying Onion Bloom Clocks}
\label{DOBC}

The naive resolution described in \autoref*{sec:design:cbc} would require an extremely inefficient $k*n$ bits. 

In \sys, we introduce a novel logical clock construct: the Decaying Onion Bloom Clock (\clk).
\clk an improvement over $BC$s:

\begin{enumerate}
    \item \clk probabilistically determines causality between objects with a depth difference of at most $k$.
    \item \clk addresses the issues described in \autoref*{sec:design:cbc}, whilst utilizing less space.
    \clk achieves this by keeping a finer grain memory of recent state transitions; This is opposed to distant state transitions, where its view is compressed to produce a coarser grained expression. 
    To provide indefinite utility across any number of state transitions, \clk eventually forgets states that are too distant. 
    Eventually, states that are too distant are forgotten.
    \item A sub-function that allows \clk{}s of different depths to be \textit{merged}. The casuality utility is maintained with regard to any of it its ancestors. 
\end{enumerate}

In this section we will describe the base \clk protocol.

\lijl{Need more intuition and high-level idea before diving into details.}

We generalize the Counting bloom filter construct to variable-sized Bloom filters ($VBF_i$), where each of its $n$ indices are stored with exactly $i$ bits.
The \clk also consists of $|L|$ layers, each layer $l^i$ stores a pre-determined amount $|l^i|$ of $VBF_{j^{i}}$s, where $j^{i}$ is the size an index for each $VBF$ at layer $l^i$ and $j^{i+1} > j^i$.
For the sake of simplicity, let's assume $j^{i}\cong i$.
Each $VBF_i$ in a layer is ordered from $l^i_1,\cdots,l^i_{|l^i|}$.

\begin{figure}[h!]
    \centering
    \includegraphics[width=.4\textwidth]{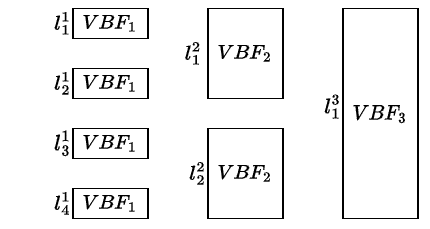}
    \caption{An illustration of how a \clk will look like with the setting: $j^{i}\cong i, |L| = 3, |l^1| = 4, |l^2| = 2, |l^3| = 1$ }
    \label{fig:example_DOBC}
\end{figure}

Suppose for a certain execution path, it produces an ordered set of objects $\{o_0,o_1,o_2,\cdots\}$, where $\exists \mu \ in \ M: \mu(o_i=(s_i,C_i,d)) \rightarrow (o_{i+1}=(s_{i+1},C_{i+1},d+1))$.
Initially, $VBF_i$s on all layers are set to 0.
For illustration purposes, let's use the following settings:
\begin{gather*}
    |L| = 3, |l^1| = 4, |l^2| = 2, |l^3| = 1
\end{gather*}
This is illustrated with \autoref*{fig:example_DOBC}.

When $o_1$ is generated, a Bloom filter $BF_{s_1}$ is created by hashing $s_1$ with the family of $m$ distinct hash functions and setting the corresponding indices to 1.
It is important to note that $VBF_1 \cong BF$ if both contain the same number of indices.
\lijl{This is only true for $VBF_1$ right?}
\mh{Yes correct}

\begin{figure}[h!]
    \centering
    \includegraphics[width=.8\textwidth]{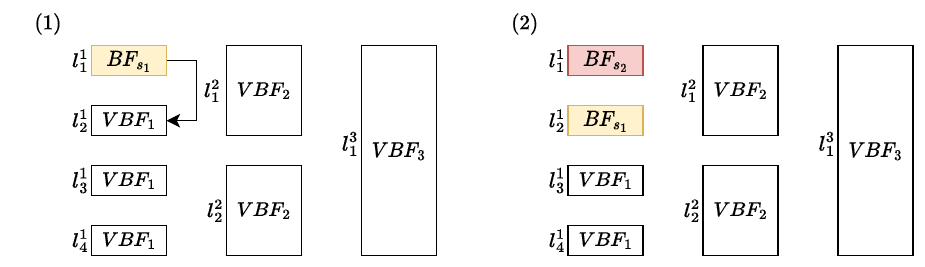}
    \caption{(1) illustrates the \clk for $o_1$, (2) illustrates the \clk for $o_2$ where $BF_{s_1}$ is moved to $l_2^1$ to make space for $BF_{s_2}$.}
    \label{fig:DOBC_example1}
\end{figure}

$BF_{s_1}$ is inserted into $l^1_1$.
When $s_2$ is reached, $BF_{s_2}$ is created and placed into $l^1_1$, and $BF_{s_1}$ is moved to the next available slot (which in this case is $l^1_2$).
\clk for $o_1$ and $o_2$ is illustrated by \autoref*{fig:DOBC_example1}.

\begin{figure}[h!]
    \centering
    \includegraphics[width=.8\textwidth]{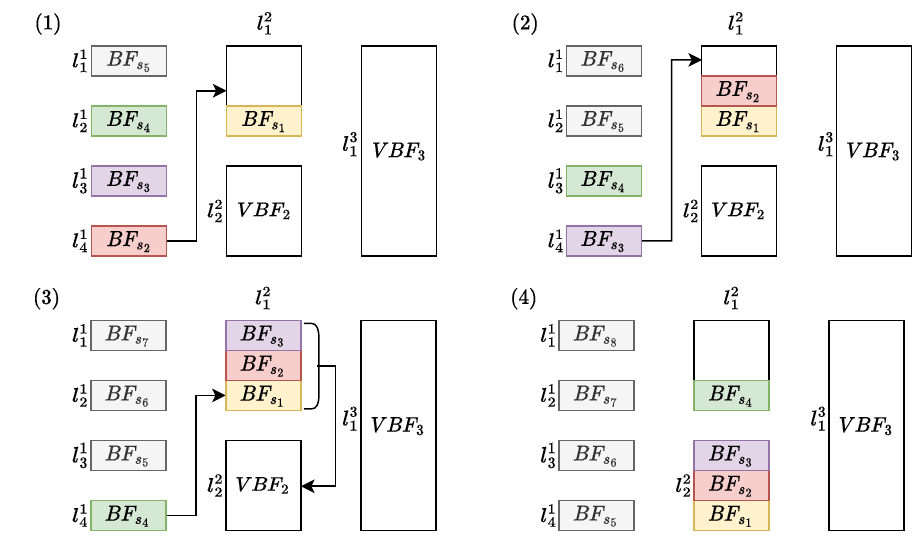}
    \caption{(1) illustrates the \clk for $o_5$, (2) \clk for $o_6$, (3) \clk for $o_7$, (4) \clk for $o_8$. }
    \label{fig:DOBC_example2}
\end{figure}

Eventually, as new objects (and states) are created, in the DOBC for a specific object ($o_4$), $BF_{s_1}$ will be at $l^1_{|l^i|}=l_4^1$.
To make space for $BF_{s_4}$, $BF_{s_1}$ instead moves to $l^2_1$.
In theory a $VBF_2$ can hold the compressed information of $2*2-1 = 3$ $VBF_1$s.
Therefore, $BF_{s_1},BF_{s_2}, BF_{s_3}$ are added together before it moves to $l_2^2$.
This is illustrated by \autoref*{fig:DOBC_example2}.
Intuitively, a $VBF_{i+1}$ in layer $l^{i+1}$ can store a multiple of $VBF_{i}$ from the previous layer ($l^{i}$).

\begin{figure}[h!]
    \centering
    \includegraphics[width=.8\textwidth]{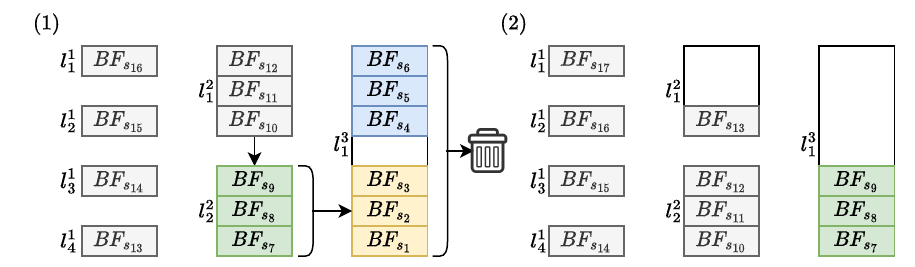}
    \caption{ (1) illustrates the \clk for $o_{16}$. (2) illustrates the \clk for $o_{17}$, where $BF{s_1},\cdots,BF_{s_6}$ is evicted. }
    \label{fig:DOBC_example3}
\end{figure}

When $l^{|L|}_{|l^i|}$ has reached the maximum capacity, and a new state is reached, for the new object, $l^{|L|}_{|l^i|}$ is \textbf{deleted} and $l^{|L|}_{|l^i-1|}$ or $l^{|L|-1}_{|l^i|}$ takes its place.
In the context of our example, $BF{s_1},\cdots,BF_{s_6}$ is evicted from $l^3_1$ and $BF_{s_7},\cdots,BF_{s_9}$ takes its space.
This is illustrated with \autoref*{fig:DOBC_example3}.

In our example, the \clk keeps track of an average of $k=13.5$ states, but it is clear to see that its ability to compare any causality drops every 4 objects.

It is clear to see that there exists some wastage in utilization with the example \clk, as a $VBF_3$ can in theory hold $2^3-1 = 7$ $BF$'s.
However, because it can only hold a certain multiple of $VBF$'s from the previous layer, it only holds $(2^2-1) * 2$ $BF$s.

The maximum number of $BF$'s or unique states held by $l^{|L|}_{|l^i|}$ is therefore:

\begin{gather*}
    \gamma =
    \left(
        \prod_{i=1}^{i=|L|-1}
            \left\lfloor
            \frac{2^{j^{i+1}}-1}{2^{j^i}-1}
            \right\rfloor
    \right)
     * (2^{j^1}-1)
\end{gather*}

The total number of states a \clk can hold thus ranges from $[k,k-\gamma+1]$ states, with an average of $k + \frac{\gamma+1}{2}$ states.
Where for $|L|> 1$:

\begin{gather*}
    k= |l^1|*(2^{j^1}-1) +
    \left(
        \sum_{i=2}^{i=|L|}
         \left\lfloor
         \frac{2^{j^i}-1}{2^{j^{i-1}}-1}
         \right\rfloor
         * |l^i|
    \right)
\end{gather*}

\begin{figure}[h!]
    \centering
    \includegraphics[width=.8\textwidth]{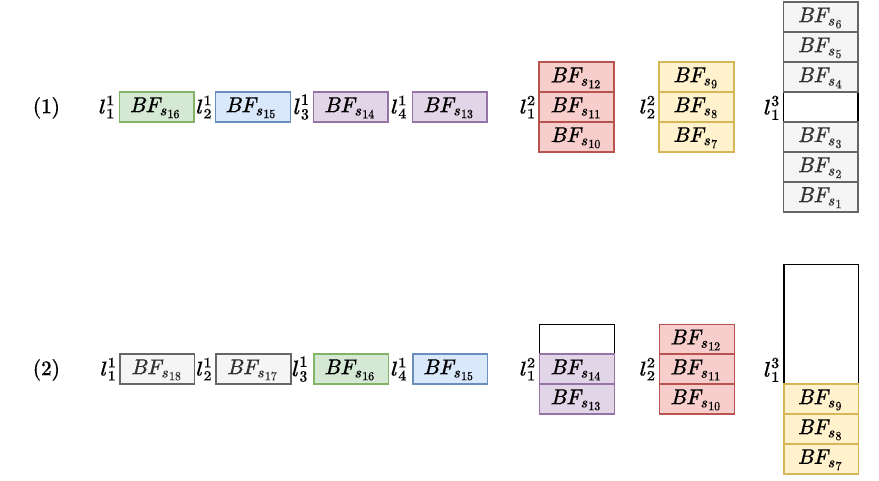}
    \caption{ (1) illustrates the \clk for $o_{16}$. (2) illustrates the \clk for $o_{18}$. To determine if $o_{18}$ is causally dependent on $o_{16}$, we simply compare the similarly highlighted parts in both clocks. If the similarly highlighted parts in (2) is greater than or equal to the ones in (1) for all similarly highlighted parts, then we conclude $o_{18} \succ o_{16}$.}
    \label{fig:DOBC_compare_example1}
\end{figure}

\subsubsection{Comparing \clk{}s}

In this section, we will go through how two different \clk{}s are compared to derive causality.
In the naive approach mentioned in \autoref*{sec:design:cbc}, it is quite obvious to see how clocks can be compared.
However, in \clk, we only keep a limited history $~k$ of states, therefore only histories of a certain range can be compared.
The greater the overlap, the lower the possibilities of false positives.
We will utilize the same setting from \autoref*{DOBC} to illustrate an example.
Suppose we have the \clk for $o_{18}$ and $o_{16}$, how we determine causality of $o_{18}$ on $o_{16}$ is illustrated with \autoref*{fig:DOBC_compare_example1}:
Since each state has differing depths, we compare different sections of its \clk{}s to draw our causality conclusion. For example, $l_3^1\in o_{18}$ should correspond to $l_1^1 \in o_{16}$. Similarly, the $l_1^2 \in o_{18}$ corresponds to addition of $l^1_3 \cap l^1_4 \in o_{16}$. Intuitively, two set of $VBF$'s are comparable between two \clk{}s if they correspond to the same depth. 
If all comparable $VBF$s in $o_{18}$ are greater than or equal to the corresponding $VBF$s in $o_{16}$, then we draw the conclusion that $o_{18} \succ o_{16}$ with some acceptable probability.

\subsubsection{Eliminating Wastage in \clk}

As briefly mentioned in \autoref*{DOBC}, certain parameters will lead to bit wastage.
This is not ideal if we wish to fully utilize every single bit in \clk.
We came up with two possible approaches to mitigate wastage.
These approaches might require overhauls to the \clk protocol:

\begin{figure}[h!]
    \centering
    \includegraphics[width=.8\textwidth]{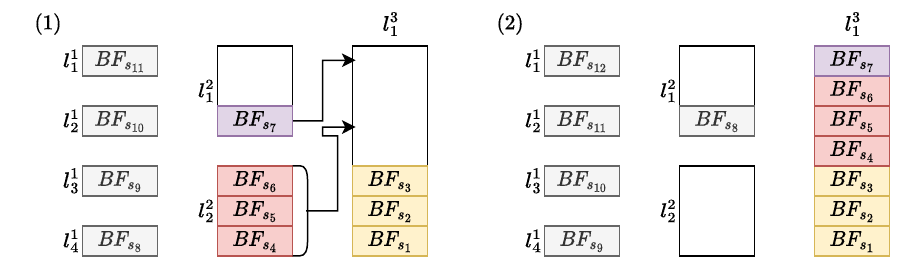}
    \caption{ (1) illustrates the \clk for $o_{11}$. (2) illustrates the \clk for $o_{12}$ with incomplete decay; Although $l^2_1$ is not completely full, it is shifted along with $1^2_2$ to completely fill $l^3_1$. }
    \label{fig:DOBC_space_example1}
\end{figure}

\begin{enumerate}
    \item \textbf{Incomplete $VBF$ decay:} We dictate that a $VBF_{j^i}$ does not necessarily have to be ``full'' before it is moved to the next layer.
    That is to say we change the ``\textit{a $VBF_{i+1}$ in layer $l^{i+1}$ can store a multiple of $VBF_{i}$ from the previous layer ($l^{i}$)}'' notion of the base \clk protocol.
    We should be able to fully utilize all the space in all layers.
    This is illustrated by \autoref*{fig:DOBC_space_example1}.
    An important note is that $VBF_{j^1}$ must be of sufficient granularity to fill all the ``gaps'' of $VBF$'s of subsequent layers.
    Therefore, $j^1 = 1$ should always work.

    \item \textbf{Perfect $VBF$ decay:} If the next layer $VBF_{j^i}$ can exactly hold a multiple of $VBF_{j^{i-1}}$s from the previous layer, then naturally, space will be fully utilized.
    More concretely, the number of $BF$s $VBF_{j^{i}}$ can hold is congruent to 0 modulo of the $BF$s $VBF_{j^{i-1}}$ can hold, given $|L|>1$.

    An example would be to have the setting: $|L| = 3, j^1 = 1, j^2 = 2, j^3 = 4$.
    No space will be wasted since:
    \begin{gather*}
        (2^2-1) \% 1 = 0 \\
        (2^4-1) \% (2^2-1) = 0
    \end{gather*}

\end{enumerate}

\subsubsection{Merging \clk{}s}

\begin{figure}[h!]
    \centering
    \hspace*{-.5cm}   
    \includegraphics[width=1.25\textwidth]{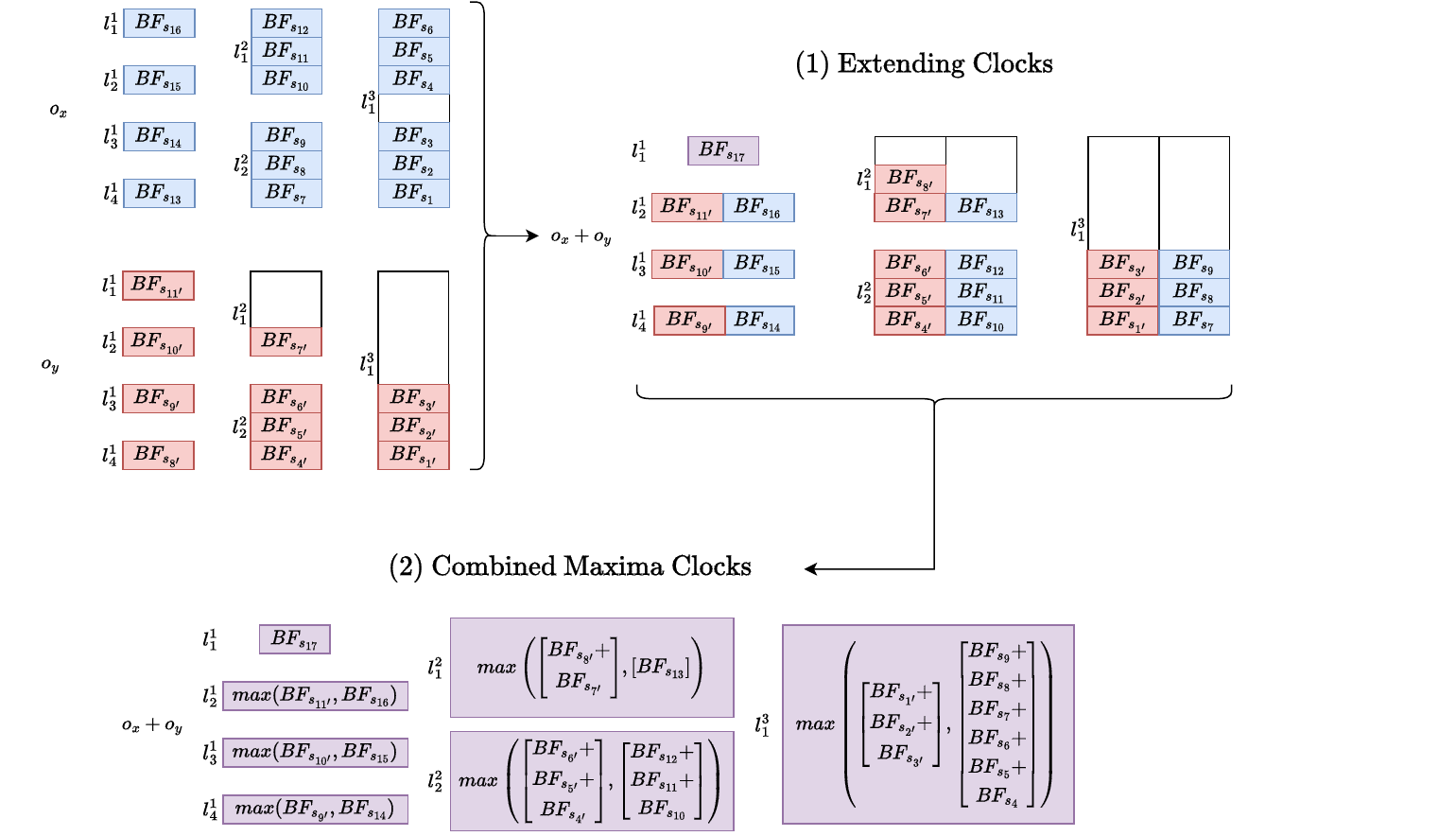}
    \caption{Utilizing the same settings from \autoref*{DOBC}: Suppose we wish to merge two objects $o_x$ and $o_y$ (1) The \textbf{Extending Clocks} solution will append the appropriate $VBF$'s to each other. (2)The \textbf{Combined Maxima Clocks} solution takes it a step further; trading false positives with space. The $VBF$'s at each position $l^i_j$ are the maximum of the indices between the $VBF$'s seen in the extending clocks solution.}
    \label{fig:DOBC_merge_example1}
\end{figure}

Suppose if two objects were to be merged, the resultant state and its descendants will have some causal relation to both parents. In \sys, merging both \clk{}s must be done elegantly. This is to ensure that the causality of future descendants can still be adequately inferred. 

\clk has two methods to merge clocks, each at opposite ends of the spectrum when it comes to utility and its trade-offs.

Since \clk only keeps track of at most $k$ objects prior to the current one. When merging two objects $o_x = (s_x,C_x,d_x), o_y = (s_y,C_y,d_y):|d_x-d_y|>k$ , the resultant object technically exists as both depth $d_y+1$ and $d_x+1$. 

\mh{Changed from last time}

The unfortunate side effect of merging is that the new merged \clk must keep track of both heights. 

\begin{figure}[h!]
    \centering  
    \includegraphics[width=.8\textwidth]{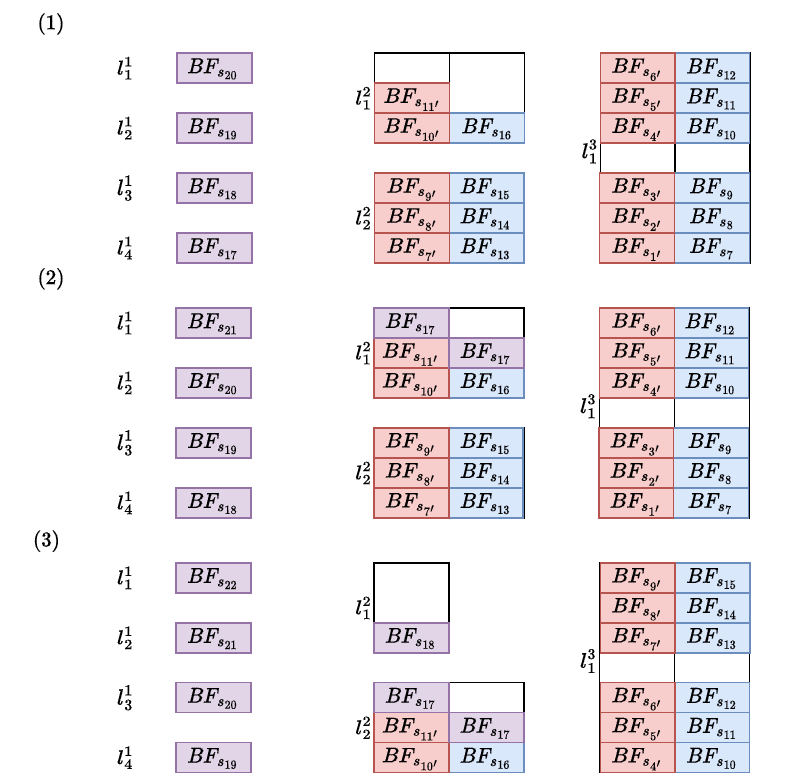}
    \caption{Since the Extending Clocks solution merges clocks at differing states of decay, decaying both clocks as is might lead to indefinite extra space utilization. To make the extra space utilization transient, we use the ``more-decayed'' clock as an anchor to decay the other ``less-decayed'' clock. As seen in (1), since the clock highlighted red is of a greater state of decay, therefore after (2), at (3) it also pre-emptively decays the blue clock. The result is a single $VBF$ at $l^2_1$. }
    \label{fig:DOBC_merge_example2}
\end{figure}

\paragraph*{\textbf{Extending Clocks: }} The Extending clocks approach is the most naive approach. As the name suggests, and illustrated in (1) of \autoref*{fig:DOBC_merge_example1}: The corresponding $VBF$'s after a merge is stored in a linked list configuration, where each $l_j^i:i=1,j>1$ is now a pair of $VBF$'s. 

Comparing \clk{}s to derive causality will now be done twice; once with the red highlighted $VBF$'s, another with the blue ones. 

This merging solution will linearly grow the size of the resultant clocks per merge. However, this increase in space is transient and will only persist for at most $k$ \texttt{mutations}. This is achieved by tweaking the decay function of \clk, by making the new merged $BF$'s decay at different rates. This is illustrated in \autoref*{fig:DOBC_merge_example2}.

\paragraph*{\textbf{Combined Maxima Clocks: }}

Instead of naively appending clocks together in the previous solution, another alternative is to take the maxima of each index of the $VBF$'s. As illustrated in \autoref*{fig:DOBC_merge_example1}, for $l^2_1$, the $VBF_2$ will have its value at each index be the maximum of that of the corresponding two $VBF$'s from $o_x$ and $o_y$. 

Intuitively, such an approach will definitely increase the false positive rate. However, its inherent benefit is that the resultant merged \clk will always remain the same size. 

Furthermore, the false positive rate can be reduced by increasing $n$ (number of indices) of the $VBF$s. 

\paragraph{\textbf{Hybrid Approach: }}
A hybrid approach will be to employ the extending clocks method for a maximum of $p$ merges. After which, older merges will be combined in the same approach outlined by the combined maxima clocks. 

This will ensure a bound on the clock size, whilst having some good utility and lower false positive rates in some cases.

\subsection{Verifiable Logical Clock Construction}
\label{sec:design:vlc}

To apply a proof system for \sys we first need to articulate the requirements we need, more precisely:

\begin{enumerate}
    \item The proof system must support a known family of functions ($M = \{\mu_1,\cdots ,\mu_i, \cdots , \mu_\omega\}, |M| = \omega$).

    \item For a particular proof corresponding to a linear execution of $n$ functions.
    The proof must assert to the validity that a non-deterministic multiset of functions of size $n$ have been applied onto the genesis state.
    That is to say $\exists m = \{m_1,\cdots, m_n\}: \forall m_i \in m, m_i \in M $ where there $\exists \pi$ that validates $s_n \leftarrow m_n(m_{n-1}\cdots(m_1(s_0)))$, where $s_0$ is the genesis state.

    \item Transparent setup: every one can join as is.

    \item Proof size and verification time must remain constant.
\end{enumerate}

At the time of writing, we will utilize the constructs introduced in Nova\cite{Kothapalli2021NovaRZ} and subsequently SuperNova\cite{Kothapalli2022SuperNovaPU}.

\subsubsection{R1CS (Rank 1 Constraint System)}
To understand Nova and by extension, SuperNova, we must first briefly explain R1CS.
R1CS is a method to quickly verify that a particular binary or arithmetic execution has been carried out properly without actually running through the execution again.
More precisely, R1CS allows the prover to generate a solution to some polynomial problem (represented by the arithmetic circuit), in which the verifier can verify in constant time.

The arithmetic circuit\footnote{We will omit details in regard to binary circuits as any binary circuit can be converted into an arithmetic circuit.} can be viewed logically as a collection of gates, where each gate has a left and right input as well as a single output.
How the gates are wired to produce the final outputs are defined by a set of 3 matrices $A, B, C \in \mathbb{F}^{q \times q}$.
The prover also generates a solution vector $Z = {W,x,1}, Z \in \mathbb{F} ^ q$, where $W$(or witness) are the intermediary outputs, $x$(or instance) are the inputs and outputs of the circuit, and $1$ is a constant.

The verifier can simply verify the execution of the whole circuit as:
\[
    (A \cdot Z) \circ (B \cdot Z) = (C \cdot Z)
\]
Where $\cdot$ denotes matrix multiplication and $\circ$ denotes the Hadamard product.

\subsubsection{What is Nova?}
Nova, is an Incrementally Verifiable Computation (IVC) algorithm that requires no trusted setup, generates constant sized proofs for any step, and guarantees constant verification time.
Nova introduced a novel method to combine two R1CS instances into a single instance.
Naively adding two R1CS instances would lead to an incorrect instance.
Therefore, Nova introduces a variant of R1CS: \emph{Relaxed R1CS} which introduces an error vector $E \in \mathbb{F}^q$, where an instance $(E,u,x)$ is satisfied if:

\[
    (A \cdot Z) \circ (B \cdot Z) = u \cdot( C \cdot Z) + E
\]

Where $Z = (W,x,u)$.
In particular, suppose there are two separate instances $Z_1 = (W_1,x_1,u_1)$ and $Z_2 = (W_2,x_2,u_2)$.
$u \leftarrow u_1 + r \cdot u_2$ and $E$ is a function of $(Z_1,Z_2,r)$.
The resulting instance or \emph{folded} instance can be simply verified by the verifier.
The intuition is that if any of the relaxed R1CS instances that were folded is invalid, then the final folded relaxed R1CS instance will be invalid as well.
Therefore, verifying the final folded relaxed R1CS instance asserts that arithmetic circuit has been executed correctly a particular number of times.

Verifier work is further reduced by the introduction of the \emph{committed relaxed R1CS} scheme.
This scheme allows the prover to utilize additively-homomorphic commitment schemes (like Pedersen commitments) to save the verifier from generating $E$ themselves.
Instead, the prover will send commitments for ${E_1,E_2,W_1,W_2}$ as well as another matrix $T$ which is a result of a function of $Z_1,Z_2$.

The verifier can additively combine the commitments to generate the committed folded instance.
The prover will then reveal the actual folded instance.
If it matches, the verifier can simply take the folded instance as is to use for verification.

The resultant proof structure for the $i^{th}$ step is a folded committed relaxed R1CS instance and witness pair asserting to the validity of executions up till step $i-1$, and a single relaxed R1CS asserting to the validity of step $i$.

The scheme is then made \emph{non-interactive} by utilizing a public coin\footnote{This can be instantiated by using a cryptographic-secure hash function} hinging on the Fiat-Shamir heuristic.

\subsubsection{Nova for a family of functions}

Traditional IVCs like Nova typically are designed for a single function.
To allow Nova to satisfy requirement (2) above, we can utilize a universal circuit.
Intuitively, a universal circuit can be visualized as a circuit made by combining $n$ sub-circuits, each representing the execution to a particular function.
The inputs of this universal circuit will then ``select'' a particular sub-circuit to execute.

The major downside to utilizing general circuits is that for any step of the execution, the whole circuit is actually still processing some input and generating some output.
This means the prover will be doing some unnecessary work. Additionally prove sizes might be larger.

SuperNova\cite{Kothapalli2022SuperNovaPU} was developed to circumvent this issue, we will elaborate this in the following text.

\subsubsection{SuperNova: Universal machines without universal circuits}
\label{sec:background:supernova}

SuperNova generalizes Nova's IVC to \emph{non-uniform} IVC, i.e., there exists a family of functions $F=\{f_1,\cdots,f_n\}$, and a control function $\varphi$ which determines the function to run at a particular step $j$.
That is to say at the $j^{th}$ step, the prover proves that $f_j, j=\varphi(W_i,x_i)$ has been correctly applied with witness instance pair $(W_i,x_i)$ to produce $x_{i+1}$.

\paragraph{\textbf{Recursive Proofs in SuperNova:}}
It is not possible to naively apply the folding scheme developed in Nova, because each function (and the corresponding circuit) is structurally heterogeneous.

A SuperNova proof therefore maintains a list of running instances $\mathtt{U}_i$, where $\mathtt{U}_i[j]$ is the folded instance of all previous invocations of $f_j$ before the $i^{th}$ step.
It also contains a corresponding list of Witnesses $\mathtt{W}_i$, as well as an instance witness pair $(\mathtt{u}_i,\mathtt{w}_i)$ that asserts to valid execution of step $i$.

Furthermore, instead of simply applying the functions within the function family as is, SuperNova instead runs the \emph{augmented} version of the function $f'_{\varphi(W_i, x_i)}$.

In essence, the augmented function does not just simply run $f_{\varphi(W_i, x_i)}(W_i,x_i)$ to output $x_{i+1}$.
It also checks that $\mathtt{U}_{i},\varphi(W_{i-1}, x_{i-1})$ are indeed produced by the prior step if it is contained in $\mathtt{u}_i$.
This asserts that checking $\mathtt{U}_{i+1}$ is the same as checking $(\mathtt{U}_{i}, \mathtt{u}_i)$.
The augmented function then folds $\mathtt{u}_i$ into $\mathtt{U}_{i}[\varphi(W_{i-1}, x_{i-1})]$, and produces $\varphi(W_i, x_i)$.

\[
((i+1,x_0,x_{i+1}),\mathtt{U}_{i+1},\varphi(W_i, x_i)) \leftarrow f'_{\varphi(W_i, x_i)} (\mathtt{U}_{i},\mathtt{u}_i,   \varphi(W_{i-1}, x_{i-1}), (i,x_0, x_i, W_i))
\]
Resulting in the witness pair $(\mathtt{u}_{i+1},\mathtt{w}_{i+1})$.

Intuitively, verifying $\mathtt{U}_{i+1}$ is equivalent to verifying the prior $i$ steps.
$\mathtt{u}_{i+1}$ asserts the $i+1$ step.
Therefore, the proof for step $i$ can be expressed as:
\[
    \Pi_{i} = ((\mathtt{U}_i, \mathtt{W}_i),(\mathtt{u}_i, \mathtt{w}_i))
\]

Additional details on SuperNova proofs using committed relaxed R1CS instances can be found in the original paper.

We aim to utilize the Non-Uniform Incrementally Verifiable Computation scheme described above for \sys.
As such, each function in $F$ corresponds injectively to a specific mutate function in $M$.
The depth value $d$ simply corresponds to $i$ in each instance $\mathtt{u}_i$.
The object state and clock value corresponds to $x_i$.

\section{Use Cases}
\label{sec:cases}
So far, we have discussed the high-level properties of \sys, and a concrete design of \sys using \clk and verifiable computation.
In this section, we describe a few use cases of \sys.

\subsection{Weakly Consistent Data Store}
\label{sec:cases:kstore}

Prior systems~\cite{dynamo} have built data stores with eventual consistency using logical clocks.
Similarly, we leverage \sys to build a weakly consistent decentralized data storage, \kstore.
\kstore implements a key-value storage interface.
Each unique \emph{key} is mapped to an arbitrarily-sized value.
\kstore is fully decentralized and permissionless.
Any node can join and leave the system at any time.
Compared to its strongly consistent counterparts~\cite{bitcoin,ethereum}, \kstore offers higher efficiency, scalability, and availability.

\kstore provides \emph{eventual consistency}: if no further writes are applied to a key, eventually all nodes observe the same value mapped to the key.
Each \kstore node maintains a subset of the keys in the key-space.
We use a distributed hash table (DHT)~\cite{chash,chord,kademlia} with virtual nodes for key-space partitioning and request routing.
For fault tolerance, each key is stored on $R$ virtual nodes closest to the key hash on the hash ring, where $R$ is a configurable parameter.
This set of virtual nodes is called the \emph{replica group} for the key.
Higher $R$ offers stronger fault tolerance, but results in longer update latency and higher storage overhead.

\kstore{}'s storage API exposes three external operations, \code{Get}, \code{Insert}, and \code{Update}.
\code{Get} takes a key and returns the value mapped to the key.
\code{Update} maps a new value to an existing key.
\code{Insert} creates a new key into \kstore with an initial mapped value.
\code{Insert} also takes an optional user-defined \code{Merge} function.
The \code{Merge} function takes a set of values as input and outputs a single value.
For instance, a \code{Merge} function for numerical value types could be \code{maximum}, and \code{union} for set value types.

When a client invokes \code{Insert}, the request is routed to one of the $R$ responsible virtual nodes on the DHT.
The client can choose any reachable nodes, $L$, in the replica group.
Upon receiving the \code{Insert} request, $L$ invokes \code{create} to generate an object, with the genesis state set to the value in the request.
The object state also includes the \code{Merge} function in the request.
$L$ then forwards the generated object to the remaining nodes in the replica group.
Each node in the group verifies the validity of the object using the attached proof (\autoref{sec:design:vlc}) and stores it locally.

When a client invokes \code{Update} on a key, the request is similarly routed to one of the $R$ responsible virtual nodes, $L$.
Note that $L$ does not need to be the same node that creates the object.
$L$ then invokes \code{mutate} with the locally stored object $o_l$ as input, \ie, ${mutate(o_l)} \rightarrow o_l'$.
The output object state is set to the value in the \code{Update} request.
$L$ then forwards $o_l'$ to the other nodes in the replica group.
When a replica node receives $o_l'$, it uses \sys to determine the causality relation between $o_l'$ and its locally stored object $o$.
If $o_l' \prec o$, the node ignores the object.
If $o \prec o_l'$, the node replaces the local object with $o_l'$.
Otherwise, $o \parallel o_l'$ and the node invokes ${mutate(o_l', o)} \rightarrow o'$, and stores the new object $o'$.
When invoking \code{mutate}, the node applies the \code{Merge} function stored in the object.

When a client invokes \code{Get}, it simply routes the request to any of the $R$ nodes in the replica group.
The node returns the object if it is stored locally.
The client iterates through the replica group until the object is found.

\subsection{Anti-Censorship Decentralized Social Network}
\label{sec:cases:ksn}

The second use case is a decentralized social network, \ksn, which we built atop \sys.
In \ksn, users (represented by a private/public key pair) publish posts (\eg, short text, blogs, and photos) to the network, and subscribe to other users to receive their posted content.
Users can also react and comment on posted content, both of which are fetched alongside the content.

\ksn stores the status and all published content of a user in a \sys object with the type \uobj.
\footnote{For simplicity, we store both the metadata and the content in the \sys object.
An optimized implementation can store content separately, and only saves content hashes, which can be used as pointers, in the \sys object.}
All posts are signed by the publishing user.
\ksn defines a \code{Update} and a \code{Merge} function.
\code{Update} takes a \uobj and produces a new \uobj with the newly published posts added to it.
\code{Merge} takes multiple \uobj{}s for the same user and merge their content to produce a new \uobj.
\lijl{Probably need to define the semantics of \code{Merge}, \eg, ignores old state in causal dependency chain.}
To read the posts of a user, a subscribed client simply fetches the corresponding \uobj.
We omit the exact format of \uobj and detailed implementation of \code{Update} and \code{Merge}.

\ksn uses a DHT~\cite{chash,chord,kademlia} for content routing.
Each user \uobj is mapped to $R$ nodes closest to its public key hash on the hash ring, with $R$ a configurable parameter.
Similar to the data store, these $R$ nodes are called the replica group of the \uobj.
To publish a post, the user first fetches its own \uobj from any of nodes in the replica group.
It can optionally cache the \uobj to avoid subsequent fetches.
If the \uobj is not available, the user creates a new \uobj by calling \code{create()}.
The user then applies \code{mutate()} on the \uobj with the \code{Update} function to add the post, and sends the resulting \uobj to all nodes in the replica group.
When a node receives a \uobj, it verifies the validity of the object (\eg, \uobj contains the correct signatures from the user) and the clock, and applies $\code{mutate(\uobj, \uobj{}')}$ with the \code{Merge()} function, where \uobj{}' is the currently stored user object.

To subscribe to another user, the client sends a \code{Subscribe} message to all $R$ nodes in the replica group of the target user.
The replicas records this subscription.
Once a replica node receives a \uobj generated by the target user, it sends a notification to the subscribed client, who then fetches \uobj from the replica.
The client verifies the validity of the object before accepting it.
Due to asynchrony and network partitions, it may receive stale or diverged \uobj{}s.
To address this issue, the client stores a \uobj for each subscribed user.
When it received a \uobj{}' from a replica, it applies $\code{mutate(\uobj, \uobj{}')}$ with the \code{Merge} function to update the object.
 
\section{Conclusion}
\label{sec:concl}
In this work, we design a new logical clock system, \sys.
\sys addresses key limitations of prior logical clock constructs.
It scales perfectly in a decentralized network with dynamic membership, and tolerates Byzantine behaviors regardless of the proportion of adversaries.
\sys achieves the above strong properties by introducing a novel logical clock structure, the Decaying Onion Bloom Clock (\clk).
It additionally applies non-uniform IVC to ensure independently verifiable construction of \clk even in the presence of Byzantine behaviors.
To showcase the capability of verifiable causality enabled by \sys, we have built a weakly consistent key-value store and an anti-censorship social network using \sys.
 
\bibliography{paper}

\begin{thebibliography}{10}

\bibitem{megastore}
J.~Baker, C.~Bond, J.~C. Corbett, J.~J. Furman, A.~Khorlin, J.~Larson, J.-M.
  Leon, Y.~Li, A.~Lloyd, and V.~Yushprakh.
\newblock Megastore: {Providing} {Scalable}, {Highly} {Available} {Storage} for
  {Interactive} {Services}.
\newblock In {\em Proceedings of the {Conference} on {Innovative} {Data} system
  {Research}}, {CIDR} '11, pages 223--234, Asilomar, California, 2011.

\bibitem{snarks}
N.~Bitansky, R.~Canetti, A.~Chiesa, and E.~Tromer.
\newblock From extractable collision resistance to succinct non-interactive
  arguments of knowledge, and back again.
\newblock In {\em Proceedings of the 3rd Innovations in Theoretical Computer
  Science Conference}, ITCS '12, page 326–349, New York, NY, USA, 2012.
  Association for Computing Machinery.

\bibitem{bloom-filter}
B.~H. Bloom.
\newblock Space/time trade-offs in hash coding with allowable errors.
\newblock {\em Commun. ACM}, 13(7):422–426, jul 1970.

\bibitem{Bowe2020RecursivePC}
S.~Bowe, J.~Grigg, and D.~Hopwood.
\newblock Recursive proof composition without a trusted setup.
\newblock 2020.

\bibitem{chubby}
M.~Burrows.
\newblock The {Chubby} {Lock} {Service} for {Loosely}-{Coupled} {Distributed}
  {Systems}.
\newblock In {\em Proceedings of the 7th {Symposium} on {Operating} {Systems}
  {Design} and {Implementation}}, {OSDI} '06, pages 335--350, Seattle,
  Washington, 2006. USENIX Association.

\bibitem{ethereum}
V.~Buterin.
\newblock A {Next}-{Generation} {Smart} {Contract} and {Decentralized}
  {Application} {Platform}.
\newblock 2014.

\bibitem{bigtable}
F.~Chang, J.~Dean, S.~Ghemawat, W.~C. Hsieh, D.~A. Wallach, M.~Burrows,
  T.~Chandra, A.~Fikes, and R.~E. Gruber.
\newblock Bigtable: {A} {Distributed} {Storage} {System} for {Structured}
  {Data}.
\newblock In {\em Proceedings of the 7th {Symposium} on {Operating} {Systems}
  {Design} and {Implementation}}, {OSDI} '06, pages 205--218, Seattle,
  Washington, 2006. USENIX Association.

\bibitem{pcd}
A.~Chiesa and E.~Tromer.
\newblock Proof-carrying data and hearsay arguments from signature cards.
\newblock In A.~C. Yao, editor, {\em Innovations in Computer Science - {ICS}
  2010, Tsinghua University, Beijing, China, January 5-7, 2010. Proceedings},
  pages 310--331. Tsinghua University Press, 2010.

\bibitem{spanner}
J.~C. Corbett, J.~Dean, M.~Epstein, A.~Fikes, C.~Frost, J.~J. Furman,
  S.~Ghemawat, A.~Gubarev, C.~Heiser, P.~Hochschild, W.~Hsieh, S.~Kanthak,
  E.~Kogan, H.~Li, A.~Lloyd, S.~Melnik, D.~Mwaura, D.~Nagle, S.~Quinlan,
  R.~Rao, L.~Rolig, Y.~Saito, M.~Szymaniak, C.~Taylor, R.~Wang, and
  D.~Woodford.
\newblock Spanner: {Google}'s {Globally}-{Distributed} {Database}.
\newblock In {\em Proceedings of the 10th {USENIX} {Conference} on {Operating}
  {Systems} {Design} and {Implementation}}, {OSDI} '12, pages 251--264,
  Hollywood, CA, USA, 2012. USENIX Association.

\bibitem{dynamo}
G.~DeCandia, D.~Hastorun, M.~Jampani, G.~Kakulapati, A.~Lakshman, A.~Pilchin,
  S.~Sivasubramanian, P.~Vosshall, and W.~Vogels.
\newblock Dynamo: {Amazon}'s {Highly} {Available} {Key}-{Value} {Store}.
\newblock In {\em Proceedings of {Twenty}-{First} {ACM} {SIGOPS} {Symposium} on
  {Operating} {Systems} {Principles}}, {SOSP} '07, pages 205--220, Stevenson,
  Washington, USA, 2007. Association for Computing Machinery.

\bibitem{gfs}
S.~Ghemawat, H.~Gobioff, and S.-T. Leung.
\newblock The {Google} {File} {System}.
\newblock In {\em Proceedings of the {Nineteenth} {ACM} {Symposium} on
  {Operating} {Systems} {Principles}}, {SOSP} ’03, pages 29--43, Bolton
  Landing, NY, USA, 2003. Association for Computing Machinery.

\bibitem{lockmanager}
A.~Hastings.
\newblock Distributed lock management in a transaction processing environment.
\newblock In {\em Proceedings Ninth Symposium on Reliable Distributed Systems},
  pages 22--31, 1990.

\bibitem{chash}
D.~Karger, E.~Lehman, T.~Leighton, R.~Panigrahy, M.~Levine, and D.~Lewin.
\newblock Consistent {Hashing} and {Random} {Trees}: {Distributed} {Caching}
  {Protocols} for {Relieving} {Hot} {Spots} on the {World} {Wide} {Web}.
\newblock In {\em Proceedings of the {Twenty}-{Ninth} {Annual} {ACM}
  {Symposium} on {Theory} of {Computing}}, {STOC} '97, pages 654--663, El Paso,
  Texas, USA, 1997. Association for Computing Machinery.

\bibitem{Kothapalli2022SuperNovaPU}
A.~Kothapalli and S.~T.~V. Setty.
\newblock Supernova: Proving universal machine executions without universal
  circuits.
\newblock {\em IACR Cryptol. ePrint Arch.}, 2022:1758, 2022.

\bibitem{Kothapalli2021NovaRZ}
A.~Kothapalli, S.~T.~V. Setty, and I.~Tzialla.
\newblock Nova: Recursive zero-knowledge arguments from folding schemes.
\newblock In {\em IACR Cryptology ePrint Archive}, 2021.

\bibitem{lamport-clock}
L.~Lamport.
\newblock Time, {Clocks}, and the {Ordering} of {Events} in a {Distributed}
  {System}.
\newblock {\em Commun. ACM}, 21(7), July 1978.

\bibitem{paxos}
L.~Lamport.
\newblock Paxos {Made} {Simple}.
\newblock {\em ACM SIGACT News}, 32(4):51--58, Dec. 2001.

\bibitem{byzantine}
L.~Lamport, R.~Shostak, and M.~Pease.
\newblock The {Byzantine} {Generals} {Problem}.
\newblock {\em ACM Trans. Program. Lang. Syst.}, 4(3):382--401, July 1982.

\bibitem{pegasus}
J.~Li, J.~Nelson, E.~Michael, X.~Jin, and D.~R.~K. Ports.
\newblock Pegasus: {Tolerating} {Skewed} {Workloads} in {Distributed} {Storage}
  with {In}-{Network} {Coherence} {Directories}.
\newblock In {\em 14th {USENIX} {Symposium} on {Operating} {Systems} {Design}
  and {Implementation}}, {OSDI} '20, pages 387--406. USENIX Association, Nov.
  2020.

\bibitem{tree-clock}
U.~Mathur, A.~Pavlogiannis, H.~C. Tun\c{c}, and M.~Viswanathan.
\newblock A tree clock data structure for causal orderings in concurrent
  executions.
\newblock In {\em Proceedings of the 27th ACM International Conference on
  Architectural Support for Programming Languages and Operating Systems},
  ASPLOS '22, page 710–725, New York, NY, USA, 2022. Association for
  Computing Machinery.

\bibitem{kademlia}
P.~Maymounkov and D.~Mazières.
\newblock Kademlia: {A} {Peer}-to-{Peer} {Information} {System} {Based} on the
  {XOR} {Metric}.
\newblock In P.~Druschel, F.~Kaashoek, and A.~Rowstron, editors, {\em
  Peer-to-{Peer} {Systems}}, {IPTPS} '02, pages 53--65, Berlin, Heidelberg,
  2002. Springer Berlin Heidelberg.

\bibitem{bitcoin}
S.~Nakamoto.
\newblock Bitcoin: {A} peer-to-peer electronic cash system.
\newblock 2009.

\bibitem{vr}
B.~M. Oki and B.~H. Liskov.
\newblock Viewstamped {Replication}: {A} {New} {Primary} {Copy} {Method} to
  {Support} {Highly}-{Available} {Distributed} {Systems}.
\newblock In {\em Proceedings of the {Seventh} {Annual} {ACM} {Symposium} on
  {Principles} of {Distributed} {Computing}}, {PODC} '88, pages 8--17, Toronto,
  Ontario, Canada, 1988. Association for Computing Machinery.

\bibitem{raft}
D.~Ongaro and J.~Ousterhout.
\newblock In {Search} of an {Understandable} {Consensus} {Algorithm}.
\newblock In {\em Proceedings of the 2014 {USENIX} {Conference} on {USENIX}
  {Annual} {Technical} {Conference}}, {USENIX} {ATC} '14, pages 305--320,
  Philadelphia, PA, 2014. USENIX Association.

\bibitem{lsmtree}
P.~O’Neil, E.~Cheng, D.~Gawlick, and E.~O’Neil.
\newblock The log-structured merge-tree (lsm-tree).
\newblock {\em Acta Inf.}, 33(4):351–385, jun 1996.

\bibitem{bloom-clock}
L.~Ramabaja.
\newblock The bloom clock.
\newblock arXiv 1905.13064, 2019.

\bibitem{smr}
F.~B. Schneider.
\newblock Implementing {Fault}-{Tolerant} {Services} {Using} the {State}
  {Machine} {Approach}: {A} {Tutorial}.
\newblock {\em ACM Comput. Surv.}, 22(4):299--319, Dec. 1990.

\bibitem{vector-clock}
R.~Schwarz and F.~Mattern.
\newblock Detecting causal relationships in distributed computations: {In}
  search of the holy grail.
\newblock {\em Distributed Computing}, 7(3):149--174, Mar. 1994.

\bibitem{chord}
I.~Stoica, R.~Morris, D.~Karger, M.~F. Kaashoek, and H.~Balakrishnan.
\newblock Chord: {A} {Scalable} {Peer}-to-{Peer} {Lookup} {Service} for
  {Internet} {Applications}.
\newblock In {\em Proceedings of the 2001 {Conference} on {Applications},
  {Technologies}, {Architectures}, and {Protocols} for {Computer}
  {Communications}}, {SIGCOMM} '01, pages 149--160, San Diego, California, USA,
  2001. Association for Computing Machinery.

\bibitem{bayou}
D.~B. Terry, M.~M. Theimer, K.~Petersen, A.~J. Demers, M.~J. Spreitzer, and
  C.~H. Hauser.
\newblock Managing update conflicts in bayou, a weakly connected replicated
  storage system.
\newblock In {\em Proceedings of the Fifteenth ACM Symposium on Operating
  Systems Principles}, SOSP '95, page 172–182, New York, NY, USA, 1995.
  Association for Computing Machinery.

\bibitem{calvin}
A.~Thomson, T.~Diamond, S.-C. Weng, K.~Ren, P.~Shao, and D.~J. Abadi.
\newblock Calvin: {Fast} {Distributed} {Transactions} for {Partitioned}
  {Database} {Systems}.
\newblock In {\em Proceedings of the 2012 {ACM} {SIGMOD} {International}
  {Conference} on {Management} of {Data}}, {SIGMOD} '12, pages 1--12,
  Scottsdale, Arizona, USA, 2012. Association for Computing Machinery.

\bibitem{plausible-clock}
F.~J. Torres-Rojas and M.~Ahamad.
\newblock Plausible {Clocks}: {Constant} {Size} {Logical} {Clocks} for
  {Distributed} {Systems}.
\newblock {\em Distrib. Comput.}, 12(4):179--195, Sept. 1999.
\newblock Place: Berlin, Heidelberg Publisher: Springer-Verlag.

\bibitem{ivc}
P.~Valiant.
\newblock Incrementally verifiable computation or proofs of knowledge imply
  time/space efficiency.
\newblock In {\em Proceedings of the 5th Conference on Theory of Cryptography},
  TCC'08, page 1–18, Berlin, Heidelberg, 2008. Springer-Verlag.

\bibitem{ceph}
S.~A. Weil, S.~A. Brandt, E.~L. Miller, D.~D.~E. Long, and C.~Maltzahn.
\newblock Ceph: {A} {Scalable}, {High}-{Performance} {Distributed} {File}
  {System}.
\newblock In {\em Proceedings of the 7th {Symposium} on {Operating} {Systems}
  {Design} and {Implementation}}, {OSDI} '06, pages 307--320, Seattle,
  Washington, 2006. USENIX Association.

\bibitem{netlock}
Z.~Yu, Y.~Zhang, V.~Braverman, M.~Chowdhury, and X.~Jin.
\newblock Netlock: Fast, centralized lock management using programmable
  switches.
\newblock In {\em Proceedings of the Annual Conference of the ACM Special
  Interest Group on Data Communication on the Applications, Technologies,
  Architectures, and Protocols for Computer Communication}, SIGCOMM '20, page
  126–138, New York, NY, USA, 2020. Association for Computing Machinery.

\end{thebibliography}

\end{document}